\documentclass[aps,pre,twocolumn,unsortedaddress,showpacs,floatfix,10pt,letterpaper]{revtex4-1}

\usepackage{graphicx}    
\usepackage{dcolumn}     
\usepackage{bm}          
\usepackage{longtable}   
\usepackage{color}       
\usepackage{multirow}
\usepackage{enumitem}
\usepackage{units}
\usepackage{url}

\begin{document}

\title{Generic folding and transition hierarchies for surface adsorption of HP lattice model proteins}

\author{Ying Wai Li}
\email[]{ywli@physast.uga.edu}
\affiliation{Center for Simulational Physics, University of Georgia, Athens, Georgia 30602, U.S.A.}
\altaffiliation{Present address: National Center for Computational Sciences, Oak Ridge National Laboratory, Oak Ridge, Tennessee 37831, U.S.A.}

\author{Thomas W\"ust}
\email[]{thomas.wuest@wsl.ch}
\affiliation{Swiss Federal Research Institute WSL, Z\"urcherstrasse 111, CH-8903 Birmensdorf, Switzerland}

\author{David P. Landau}
\email[]{dlandau@physast.uga.edu}
\affiliation{Center for Simulational Physics, University of Georgia, Athens, Georgia 30602, U.S.A.}

\date{\today}

\begin{abstract}

The thermodynamic behavior and structural properties of
hydrophobic-polar (HP) lattice proteins interacting with attractive
surfaces are studied by means of Wang-Landau sampling. Three benchmark
HP sequences (48mer, 67mer, and 103mer) are considered with different
types of surfaces, each of which attracting either all monomers, only
hydrophobic (H) monomers, or only polar (P) monomers,
respectively. The diversity of folding behavior in dependence of
surface strength is discussed. Analyzing the combined patterns of
various structural observables, such as e.g., the derivatives of the
numbers of surface contacts, together with the specific heat, we are
able to identify generic categories of folding and transition
hierarchies. We also infer a connection between these transition
categories and the relative surface strengths, i.e., the ratio of the
surface attractive strength to the inter-chain attraction among H
monomers. The validity of our proposed classification scheme is
reinforced by the analysis of additional benchmark sequences. We thus
believe that the folding hierarchies and identification scheme are
generic for HP proteins interacting with attractive surfaces,
regardless of chain length, sequence, or surface attraction.

\end{abstract}

\pacs{05.10.-a, 82.35.Gh, 87.15.ak, 87.15.Cc}

\keywords{}

\maketitle


\section{Introduction \label{intro}}

Protein adsorption on solid surfaces has attracted strong research
interest recently for its numerous applications in nanotechnology and
biomaterials \cite{hlady1,gray}. The study of protein functions in
experiments often involves the immobilization of proteins
\cite{macbeath,phizicky}. Adhesion of proteins on solid substrates
including metals, semiconductors, carbon or silica etc., enables the
synthesis of new materials for biosensors or electronic devices
\cite{brown,whaley,goede,bachmann1}. It is also the key to reveal the
principles of many biological processes and causes of diseases, e.g.,
when integrating implanted materials with body tissues \cite{andrade,
horbett}. Understanding how the functions and conformations of a
protein are affected by adsorption and desorption is an important
topic in protein drug delivery \cite{gombotz,pinholt}.

It is known that configurational changes of protein molecules upon
surface adsorption depend on both the protein's properties (e.g.
sequence, size, thermodynamic stability, etc.) and the surface
properties (e.g. materials, polarity, surface roughness, etc.); but
how large these changes are and where in the protein molecules they
occur remain puzzles to be solved \cite{hlady1,rabe}. Enormous
efforts have been dedicated to unveil the mysteries in protein
folding and adsorption mainly by experimental approaches \cite{ramsden,
hlady2}. Nevertheless, the fact that only the ``final product'' can
be obtained and studied in an experiment makes for slow progress in
understanding the dynamics and folding processes. From another point
of view, the diversity of possible protein sequences and sophisticated
interactions among amino acids also complicate theoretical structural
prediction in protein folding - not to mention when the protein
interacts with solvent molecules or a substrate where an extra level
of complexity enters.

With the advances in computer power, numerical simulation has become a
promising way to shed more light on the general problem. The study of
simplified, coarse-grained protein models in conjunction with Monte
Carlo simulations, being able to efficiently explore large
conformational phase spaces \cite{landau,ganazzoli}, has been a
particularly successful approach in understanding the principles
behind protein folding and adsorption.  Among these
  minimalist models, the hydrophobic-polar (HP) model
  introduced by Dill et al. \cite{dill,lau} has been a particularly
  active, yet difficult, research subject in recent years. The
  interest in the HP model is twofold: (i) Despite its known
  limitations \cite{chan,losrios}, the model captures some of the most
  important qualitative features which drive the folding of proteins
  and characterize their native states. It thus laid a basis to
  systematically study many problems in protein folding by means of
  computer simulations. (ii) The ground state search for an HP
  sequence is an NP-complete problem \cite{berger,crescenzi}, and the
  rough free energy landscapes cause traditional Monte Carlo methods,
  e.g. Metropolis sampling \cite{metropolis}, to fail in the low
  temperature regime \cite{li3}. Therefore, the HP model is a
  well-suited testing ground for a number of emerging numerical
  methods \cite{unger,frauenkron,hsu,hsu2,zhang,
  iba,bachmann2,prellberg,kou}. In addition, the thermodynamic
  behavior of different HP sequences can vary noticeably, even for the
  same chain length \cite{bachmann2}. Consequently, finite-size
  scaling cannot be applied to a systematic study of the effect of
  system size, unlike many other models in statistical physics. Thus,
  our understanding of the general behavior of the model is still
  incomplete and the simulation of the HP model remains a challenging
  computational problem in statistical physics.

Various approaches have been undertaken to better understand the
energy landscapes, thermodynamics and conformational transitions of HP
proteins adsorbing on an attracting surface \cite{liu,liu2,castells1,
  castells2,bachmann3,swetnam}. In this work, our intent is to
identify generic thermodynamic and structural behavior of protein
adsorption from a macroscopic perspective using the HP model. We have
adopted Wang-Landau sampling \cite{wls,wls2,wls3} to obtain the
density of states of a system, from which subsequent thermodynamics
can be calculated. To the best of our knowledge, it is the first
comprehensive analysis of structural transitions for protein
adsorption that integrates results from \textit{multiple} HP
sequences. Previous work by Bachmann and Janke \cite{bachmann3},
Swetnam and Allen \cite{swetnam} or Radhakrishna et
al. \cite{radhakrishna} only studied the conformational pseudo-phases
based on individual benchmark HP sequences. Comparing the
thermodynamic and structural properties of \emph{multiple} sequences
allowed us to identify categories of generic transition behavior and
to draw a correspondence between these categories and the relative
interaction strengths involved. This finding is fundamental as it
implies that different HP sequences share certain general,
qualitative, characteristics in structural transformations when
brought near to an adsorbing substrate.

The article is organized as follows: Section \ref{model} describes the
model employed, Section \ref{method} explains the sampling method and details,
Section \ref{results} presents the simulation results, Section \ref{discussion}
discusses the analysis and interpretation of the results, and Section
\ref{conclusion} gives a summary and outlook of this work.


\section{Model \label{model}}

The commonly known 22 amino acids found in proteins can roughly be
classified as either hydrophobic or polar depending on the nature of their
side chains. The tendency of the non-polar residues to stay away from water
molecules has been identified as the key driving ``force'' in forming tertiary
structures. The hydrophobic-polar (HP) model \cite{dill,lau} is a coarse-grained
lattice model for proteins that captures this hydrophobic effect. In this
model, an amino acid is treated as a single monomer and it can only be
hydrophobic (H) or polar (P). A protein is thus represented by a heteropolymer
which consists of $N$ connected monomers of type H or P. An attractive
interaction exists only between a pair of non-bonded nearest neighboring H
monomers. This attraction is denoted by $\varepsilon_{HH}$ in our discussion,
and the magnitude indicates the ability of the H monomers to pull themselves
together as determined by the insolubility of the protein in an aqueous
environment. In other words, the solvent quality is intrinsically considered
by the model. Other factors like charges and acidity of amino
acids that also govern protein folding are not handled in this scope.

In addition to the internal interactions within the polymer, the binding of a
protein with an attractive substrate contributes to the total energy. We
have considered three types of surface fields in view of the setting of the
HP model: (i) a surface interacts only with H monomers (a hydrophobic surface)
with strength $\varepsilon_{SH}$, (ii) a surface interacts only with P monomers
(a polar surface) with strength $\varepsilon_{SP}$, and (iii) a surface
interacts with both H and P monomers with equal strength, i.e., $\varepsilon_{SH}
= \varepsilon_{SP} \neq 0$. The energy function of an HP sequence interacting
with a surface takes the general form:
\begin{equation}
E = -\varepsilon_{HH}n_{HH} - \varepsilon_{SH}n_{SH} - \varepsilon_{SP}n_{SP},
\label{Hamiltonian}
\end{equation}

\noindent
where $n_{HH}$ denotes the number of interacting pairs between H monomers,
$n_{SH}$ the number of surface contacts with H monomers and $n_{SP}$ the
number of surface contacts with the P monomers. Besides contributing to the
system's energy, these three quantities are also useful ``order'' parameters
that give quantitative measures of the structure of a conformation
\cite{bachmann4}. The negative signs in front of each term indicate
that it is energetically favorable when the monomers interact or come
in contact with the surface.

Our simulations were performed on a three-dimensional simple cubic lattice
with the attractive surface being the $xy$-plane placed at $z = 0$. A
non-interacting steric wall is placed at $z = N + 1$ to confine the polymer
in a way that it can contact both walls with its ends only when it is a
vertical straight chain. Periodic boundary conditions are imposed in the $x$
and $y$ directions.


\section{Method\label{method}}

\subsection{Calculation of thermodynamic quantities}

The partition function, $Z$, at a particular temperature $T$ can be expressed
in terms of the energy density of states $g(E)$:
\begin{equation}
Z = \sum_{E}{g(E) e^{-E/k_B T}},
\label{PartitionFunction}
\end{equation}

\noindent
where $E$ is the energy of the system as defined by the energy function,
$k_B$ is the Boltzmann constant, and the sum runs over all the energies
that the system can take. Since $g(E)$ does not depend on $T$,
one may calculate $Z$ at any temperature with a single estimate of $g(E)$.
All the thermodynamic quantities then follow from the knowledge of $Z$. For
example, the average energy $\left\langle E \right\rangle$ and the heat
capacity $C_V$ are calculated as:
\begin{equation}
\left\langle E \right\rangle = \frac{1}{Z}\sum_{E}{E g(E) e^{-E/k_B T}},
\label{energy}
\end{equation}

\begin{equation}
C_V = \frac{\left\langle E^2 \right\rangle - \left\langle E \right\rangle^2}{k_B T^2}.
\label{Cv}
\end{equation}
The specific heat is then defined as $C_V / N$ accordingly.


\subsection{Wang-Landau sampling and trial moves}

Wang-Landau (WL) sampling is a powerful, iterative algorithm to
estimate the density of states, $g(E)$, with high accuracy. Details of
the algorithm are described in Refs.\cite{wls,wls2,wls3}. In our
simulations, rather stringent parameters were chosen in order to
obtain accurate estimates for $g(E)$: We used a flatness criterion $p
= 0.8$ for the 48mer and $p = 0.6$ for the 67mer and 103mer. The final
modification factor was set to $\ln(f_{final}) = 10^{-8}$ in all
cases.

It has been found that two types of trial moves, pull moves
\cite{pull} and bond-rebridging moves \cite{cutjoin}, work
particularly well together with WL sampling in search of the global
energy minimum conformations and the determination of the density of
states for lattice polymers \cite{HPwls1,HPwls2,wuest2}.  The ability
of reaching low energy states allows for a more thorough survey of
conformational space, yielding a higher resolution of $g(E)$ and thus
more precise thermodynamic quantities especially in the low
temperature regime. This is of particular importance for longer chain
lengths with more complex energy landscapes \cite{wuest2}.

The two trial moves are called with different
probabilities. Bond-rebridging moves transform a polymer from one
compact state to another without uncoiling, making it more efficient
than pull moves in dealing with densely packed polymers. However,
since the energy difference resulting from a bond-rebridging move is
relatively large, its acceptance rate is rather low. This drawback is
compensated for with a higher calling ratio. In our simulations, we
used move fractions of 80\% and 20\% for bond-rebridging moves and
pull moves, respectively.

In order to fulfill detailed balance when employing pull moves, an extra factor
is added to the acceptance probability of moving from state $A$ (with energy
$E_A$) to state $B$ (with energy $E_B$):
\begin{equation}
P(A \rightarrow B) = \textnormal{min}\left(1, \frac{g(E_A)}{g(E_B)}
                     \frac{n_{B \rightarrow A}/n_B}{n_{A \rightarrow B}/n_A} \right),
\label{acceptance_prob}
\end{equation}
where $n_{A \rightarrow B}$ is the number of pull moves that transform $A$ to $B$;
$n_A$ is the number of possible pull moves which can be performed from state $A$;
$n_{B \rightarrow A}$ and $n_B$ are defined likewise. Because of reversibility
of pull moves, $n_{A \rightarrow B} = n_{B \rightarrow A}$ and the two terms cancel out in Eq. (\ref{acceptance_prob}). When
a pull move is chosen to generate a new configuration, a list of possible moves
from state $A$ is first constructed to obtain $n_A$. A move is then selected
randomly from the list, generating state $B$. $n_B$ is counted and
$P(A \rightarrow B)$ can then be calculated. This procedure is computationally
expensive (it slows down the simulation by approximately an order of magnitude);
however, it secures the reliability of our results by taking detailed balance
into careful consideration.

We adopted RANLUX as the random number generator (\texttt{gsl\textunderscore
rng\textunderscore ranlxd2}), which is recommended by the GNU Scientific
Library (GSL) for its ``reliable source of uncorrelated numbers'' and
``strongest proof of randomness'' \cite{gsl}. It uses a
lagged-Fibonacci-with-skipping algorithm \cite{luescher} with double precision
output and a long period of about $10^{171}$.

\begin{figure}[b!]
  \includegraphics[width=\columnwidth]{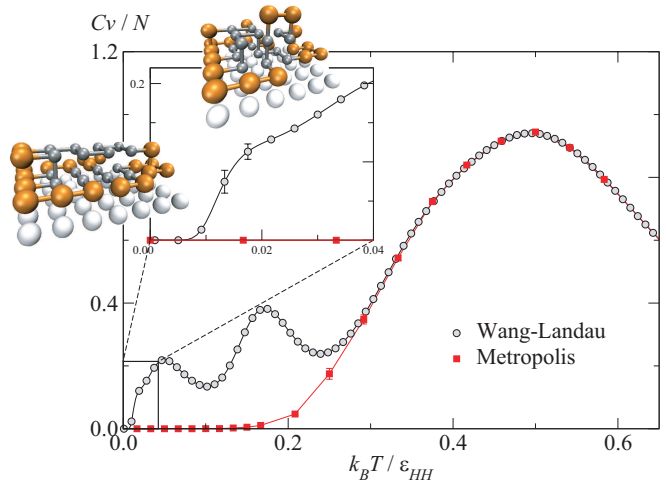}
  \caption{(Color online) \textbf{Main graph:} Specific heat of a
                 36mer interacting with a very weak attractive
                 surface as obtained by Wang-Landau and Metropolis
                 sampling. Where not shown statistical errors are
                 smaller than the data points. Comparable
                 computational effort was used for each method.
                 \textbf{Inset:} Results at extremely low
                 temperature. Typical configurations above and below
                 the shoulder are shown.}
   \label{WLvsMetro}
\end{figure}

The use of Wang-Landau sampling together with inventive trial moves is
\emph{essential} for uncovering the subtle, low temperature thermodynamics
of these systems. This is demonstrated in Fig.~\ref{WLvsMetro} which
shows the specific heat of a 36mer
(P$_3$H$_2$P$_2$H$_2$P$_5$H$_7$P$_2$H$_2$P$_4$H$_2$P$_2$HP$_2$)
interacting with a very weak attractive surface ($\varepsilon_{SH} =
\varepsilon_{SP} = 1, \varepsilon_{HH} = 12$). As seen in
the figure, all the surface related transitions below $k_BT
/ \varepsilon_{HH} \approx 0.3$ are completely inaccessible by
Metropolis sampling, even with very long runs ($10^8$ trial moves!)
incorporating pull moves and bond rebridging moves! In contrast,
Wang-Landau sampling is even able to uncover the shoulder at $k_BT /
\varepsilon_{HH} \approx 0.016$ shown in the inset (caused by a rapid
change in configurational entropy of the excitation from the ground
state to the first few excited states; see Ref. \cite{li2} for details). We,
therefore, stress that sophisticated Monte Carlo algorithms are
crucial in simulating systems with delicate low temperature
behavior.

Recently, improvements have been proposed to speed up and ensure
the convergence of WL sampling in simulating lattice polymers or proteins
\cite{swetnam,swetnam2,radhakrishna,wuest2}.


\subsection{Calculation of thermodynamics of structural observables}

Structural quantities are essential in understanding non-energetic
properties of the system. They provide information on the structures and
packing of the polymer. Apart from the quantities $n_{HH}$, $n_{SH}$ and
$n_{SP}$ introduced in Section \ref{model}, structural quantities that are
often of interest include the radius of gyration,
\begin{equation}
R_g = \sqrt{\frac{1}{N}\sum_{i=1}^N{(\vec{r}_i - \vec{r}_{cm})^2}},
\label{rg}
\end{equation}

\noindent
and the end-to-end distance,
\begin{equation}
R_{ee} = \left| \vec{r}_N - \vec{r}_1 \right|,
\label{ee}
\end{equation}

\noindent
where $\vec{r}_{cm}$ in Eq. (\ref{rg}) is the center of mass of the
configuration; $\vec{r}_i$ is the position of monomer $i$.

To obtain the thermodynamics of a structural observable $Q$, we estimated the
joint density of states, $g(E,Q)$, by multicanonical sampling \cite{berg,
berg2}. Although it is feasible to sample $g(E,Q)$ all over again using a
two-dimensional random walk in WL sampling if only one structural quantity is
required, it becomes impossible if several of them are of interest. A more
efficient way is to make use of the prior knowledge of $g(E)$ and perform a
multicanonical sampling. In this process, trial states are accepted or rejected
according to $1/g(E)$, where $g(E)$ is the one obtained previously by the
one-dimensional WL sampling and is held fixed throughout the whole production
procedure. As a new trial state is accepted, any desired structural quantity
$Q$ would be calculated and accumulated in a two-dimensional histogram, $H(E,Q)$.
The simulation is brought to an end when a sufficiently large number of Monte Carlo
steps have been performed. The joint density of states, $g(E,Q)$, is then obtained by
reweighing $H(E,Q)$:
\begin{equation}
g(E,Q) = g(E)H(E,Q).
\label{jointDOS}
\end{equation}
As such, we can obtain as many $g(E,Q)$'s for various $Q$'s as desired in a single production run.

The partition function, $Z_Q$, for observable $Q$ and its expectation value
can then be obtained as
\begin{equation}
Z_Q = \sum_{E,Q}{g(E,Q)e^{-E/k_B T}},
\label{ObservablePartitionFunction}
\end{equation}
and
\begin{equation}
\left\langle Q \right\rangle = \frac{1}{Z_Q}\sum_{E,Q}{Q g(E,Q) e^{-E/k_B T}}.
\label{observable}
\end{equation}


\section{Simulation Results \label{results}}

We have studied three benchmark HP sequences (48mer, 67mer, and
103mer) interacting with three types of surfaces (see above). The 48mer
(PHPHP$_4$HPHPHP$_2$\-HPH$_6$P$_2$H$_3$\-PH\-P$_2$HPH$_2$P$_2$HPH$_3$P$_4$H)
is seq. 9 among the ten ``Harvard sequences'' designed originally for algorithm
testing purpose \cite{48mer}, where the number of H and P monomers are exactly
the same in each sequence, i.e., 24 H monomers and 24 P monomers respectively.
The one chosen here for our simulation has the minimum ground state degeneracy
in 3D free space \cite{48mer, bachmann2}. The 67mer
(PHPH$_2$PH$_2$PH\-PP\-H$_3$P$_3$HPH$_2$PH$_2$PHP$_2$H$_3$P$_3$HPH$_2$PH$_2$PH\-P$_2$H$_3$P$_3$HP\-H$_2$P\-H$_2$\-PH\-P$_2$H$_3$P)
was first introduced to resemble $\alpha$/$\beta$-barrel in real proteins
\cite{67mer}, while the 103mer
(P$_2$H$_2$P$_5$H$_2$\-P$_2$H$_2$PH\-P$_2$\-H\-P$_7$HP$_3$\-H$_2$P\-H$_2$P$_6$\-HP$_2$HPH\-P$_2$H\-P$_5$H$_3$\-P$_4$H$_2$P\-H$_2$P$_5$H$_2$\-P$_4$\-H$_4$P\-HP$_8$\-H$_5$P$_2$HP$_2$)
was proposed to model cytochrome \textit{c} \cite{103mer}.


\subsection{Ground states and limiting behavior \label{limit}}

Table~\ref{energytable} reports the lowest energies found for these
three sequences during the estimation of $g(E)$ with different types
of attractive surfaces. To understand the ``asymptotic'' folding
behavior in the limit of a surface with infinite attractive strength,
we simulated the case where the HP chain was confined to a
two-dimensional free space. This is equivalent to restricting all
monomers of the HP chain on the surface, giving a 2D ground
state. Another limiting case is when the surface is totally absent in
a three-dimensional space. For more details on these two cases, see
\cite{li}.

\begin{table}[!htb]
\caption{Lowest energies found for the 48mer, 67mer, and 103mer interacting
         with different attractive surface types and strengths, which are
				 abbreviated in the surface labels ($A$, $H$ or $P$ stand for the surface
				 types, the numbers stand for the ratio between $\varepsilon_{SH}$
				 or $\varepsilon_{SP}$ and $\varepsilon_{HH}$). Classification of
				 transition categories are denoted by the Roman numbers in parentheses.
				 \label{energytable}}
\begin{ruledtabular}
\begin{tabular}{cccc|rc|rc|rc}
surface & $\varepsilon_{HH}$ & $\varepsilon_{SH}$ & $\varepsilon_{SP}$ &
\multicolumn{2}{c|}{48mer} & \multicolumn{2}{c|}{67mer} & \multicolumn{2}{c}{103mer} \\
\hline \\
\multicolumn{10}{l}{Free space without surface:} \\
2D  & 1      & /      & /     & \multicolumn{2}{c|}{$-21$} & \multicolumn{2}{c|}{$-29$} & \multicolumn{2}{c}{$-32$} \\
3D  & 1      & /      & /     & \multicolumn{2}{c|}{$-34$} & \multicolumn{2}{c|}{$-56$} & \multicolumn{2}{c}{$-58$} \\
\hline \\
\multicolumn{10}{l}{Surfaces attract all monomers:} \\
$A$1               & 1      & 1      & 1     & $-69$  & (I)   & $-96$  & (I)   & $-135$ & (I)    \\
$A$2               & 1      & 2      & 2     & $-117$ & (I)   & $-163$ & (I)   & /      &        \\
$A$\nicefrac{1}{2} & 2      & 1      & 1     & $-93$  & (II)  & $-132$ & (II)  & $-167$ & (I/II) \\
\hline \\
\multicolumn{10}{l}{Surfaces attract only H monomers:} \\
$H$1               & 1      & 1      & 0     & $-49$ & (II)   & $-72$  & (II)  & $-80$  & (II)   \\
$H$2               & 1      & 2      & 0     & $-73$ & (I)    & $-108$ & (I)   & /      &        \\
$H$\nicefrac{1}{2} & 2      & 1      & 0     & $-79$ & (III)  & $-118$ & (II)  & $-128$ & (III)  \\
\hline \\
\multicolumn{10}{l}{Surfaces attract only P monomers:} \\
$P$1               & 1      & 0      & 1     & $-48$ & (II)   & $-69$  & (II)  & $-100$ & (I/II) \\
$P$2               & 1      & 0      & 2     & $-71$ & (I)    & $-91$  & (I)   & /      &        \\
$P$\nicefrac{1}{2} & 2      & 0      & 1     & $-79$ & (III)  & $-123$ & (II)  & $-150$ & (II)   \\
\end{tabular}
\end{ruledtabular}
\end{table}

\begin{figure}[bth!]
  \includegraphics[width=\columnwidth]{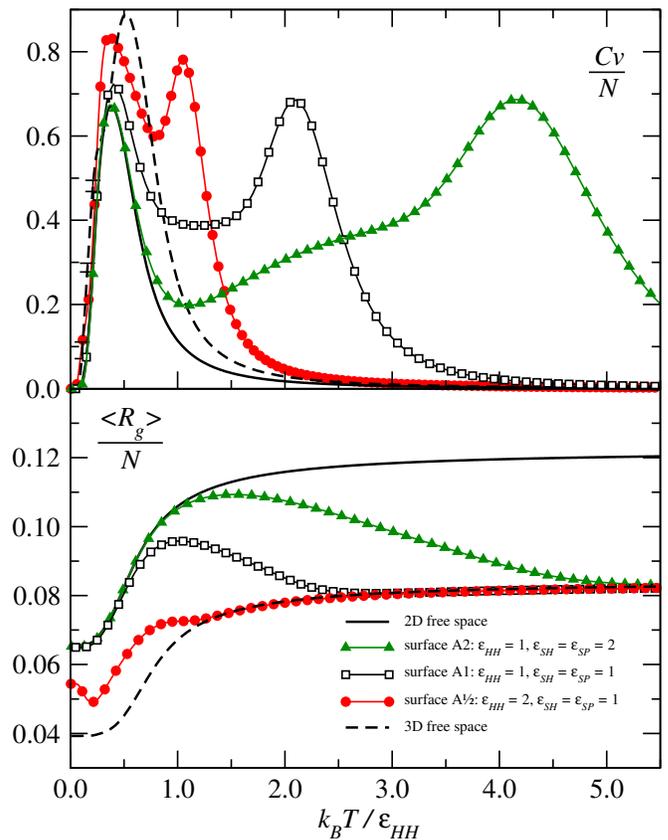}
  \caption{(Color online) \textbf{Upper panel:} Specific heat $C_V / N$
	         as a function of the effective temperature $k_BT /
					 \varepsilon_{HH}$.
					 \textbf{Lower panel:} Averaged radii of gyration per monomer,
					 $\left\langle R_g \right\rangle / N$, as a function of
					 $k_BT / \varepsilon_{HH}$, for the 48mer interacting with a surface
					 which attracts all monomers with different strengths (surfaces
					 $A$1, $A$2 and $A$\nicefrac{1}{2}). Note that $k_BT$ is scaled
					 with the internal attraction strength, $\varepsilon_{HH}$, so
					 as to compare different systems in the same energy scale. In
					 this manner, any difference in quantities comes solely from
					 the surface strengths $\varepsilon_{SH}$ and $\varepsilon_{SP}$.
					 Errors smaller than the data points are not shown.
  \label{Rg48SurfaceHP}}
\end{figure}

These two limiting cases are useful in visualizing upper and lower bounds
for thermodynamic observables and they serve as an aid to understand the
details of folding behavior. A demonstrative example is the averaged
radius of gyration per monomer, $\left\langle R_g \right\rangle / N$, shown in
Fig. \ref{Rg48SurfaceHP} for the 48mer interacting with a surface
attracting all monomers (surfaces $A$1, $A$2 and $A$\nicefrac{1}{2}). The
radii of gyration of the two limiting cases are plotted on the same figure.
Drawing a simple connection to the self-avoiding random walk on square and
cubic lattices, it is obvious that $\left\langle R_g \right\rangle$ is
largest when all the monomers are forced to sit on the surface to form
planar structures, yielding the upper bound for $\left\langle R_g \right
\rangle$. Correspondingly, $\left\langle R_g \right\rangle$ is smallest
when the HP chain is allowed to fold freely in a three-dimensional space
to form 3D structures, giving the lower bound of $\left\langle R_g \right
\rangle$. Generally speaking, when the HP chain is placed near a surface
of finite attractive strength, it remains as an extended coil at high
temperature as if the surface is absent. The radii of gyration for all
cases thus coincide with the 3D, surface-free one.

As the temperature decreases, the HP chain interacting with a stronger
attractive surface ($A$2) starts the adsorption process the earliest at
$k_BT / \varepsilon_{HH} \approx 5.0$ as its $\left\langle R_g \right\rangle$
``departs'' from the lower bound and begins to approach the upper bound.
Such an adsorption ``transition'' is clearly signaled by the peak in $C_V$
centered at $k_BT / \varepsilon_{HH} \approx 4.25$. At $k_BT /
\varepsilon_{HH} \approx 1.0$, $\left\langle R_g \right\rangle$ merges with
the upper bound signifying a complete adsorption of all monomers. The
formation of a hydrophobic core in which the number of inter-chain H-H
interactions, $n_{HH}$, is maximized, then takes place entirely on the
surface in this case until the ground state is reached at zero temperature.
This process in the low temperature regime is identical to the one in
two-dimensional free space, as indicated by the complete agreement in the
radii of gyration and the coincidence of the peak at $k_BT / \varepsilon_{HH}
\approx 0.5$ observed in $C_V$.

The thermodynamics for surface $A$1 is qualitatively similar to that
of surface $A$2 except that it requires a lower adsorption temperature.
Since the radii of gyration for both surface types end up with the same value
as the upper bound at $T = 0$, one may expect that the ground state
conformations for both systems are two-dimensional. This has been confirmed
by the number of surface contacts ($n_{SH} = n_{SP} = 24$, meaning the entire
chain is in contact with the surface) and the number of H-H interactions
($n_{HH} = 21$, which is the same as the ground state of the 2D limiting case).

While the two peaks in the specific heat of surface $A$\nicefrac{1}{2}
tend to give the impression that it has the same qualitative folding
behavior as the previous cases, the shape of the radius of gyration
clearly distinguishes it from the others, apart from showing that the
ground state is now three-dimensional. This is the first clue that the
specific heat alone does not provide a complete picture of structural
transition behavior. Indeed, the transition hierarchy of the 48mer
interacting with surface $A$\nicefrac{1}{2} is different from surfaces
$A$1 and $A$2, which can only be verified by examining other structural
parameters as we shall see in the following section.


\subsection{Identification of transition categories by canonical analysis of
specific heat and structural quantities \label{types}}

Three major transition categories were identified from the 24 systems presented
in Table \ref{energytable} by considering the combined patterns of the specific
heat $C_V / N$ and the average radius of gyration $\left\langle R_g
\right\rangle / N$.

\textbf{Category I:} $C_V$ shows two peaks, a bump between the peaks might be
                     possible, $\left\langle R_g \right\rangle$ shows a maximum
                     between these two peaks.

\textbf{Category II:} $C_V$ shows two peaks, $\left\langle R_g \right\rangle$
                      decreases upon cooling. In the very low temperature regime,
                      it might rise back up a little to form a minimum when the
                      temperature approaches zero.

\textbf{Category III:} $C_V$ shows only one peak with possible shoulders,
                       $\left\langle R_g \right\rangle$ decreases upon cooling.

The different combinations of $C_V$ and $\left\langle R_g \right\rangle$ in the
transition categories are caused by the different order of occurrence in folding
processes, which are revealed by further investigation of proper structural
parameters and their derivatives. Quantities which are particularly informative
for our systems include the derivatives of the average number of H-H interactions,
$d\left\langle n_{HH} \right\rangle / dT$, and those of the numbers of surface
contacts, $d\left\langle n_{SH} \right\rangle / dT$ and $d\left\langle n_{SP}
\right\rangle / dT$. A peak in $d\left\langle n_{HH} \right\rangle / dT$ signals
the construction of H-H interactions to form a hydrophobic core (H-core formation).
Peaks in $d\left\langle n_{SH} \right\rangle / dT$ and $d\left\langle n_{SP}
\right\rangle / dT$ provide information about the formation of surface contacts,
which is associated with the adsorption process as well as the ``flattening''
of structure due to surface attraction.

We observe that $\left\langle R_{ee} \right\rangle$ behaves quite similarly as
$\left\langle R_g \right\rangle$ but is less reliable at low temperature where mainly
compact structures are found. For this reason our analysis relies on
$\left\langle R_g \right\rangle$ rather than on $\left\langle R_{ee} \right\rangle$.


\subsubsection{Folding behavior with a strong attractive surface: Category I}

\begin{figure}[tbh!]
  \includegraphics[width=\columnwidth]{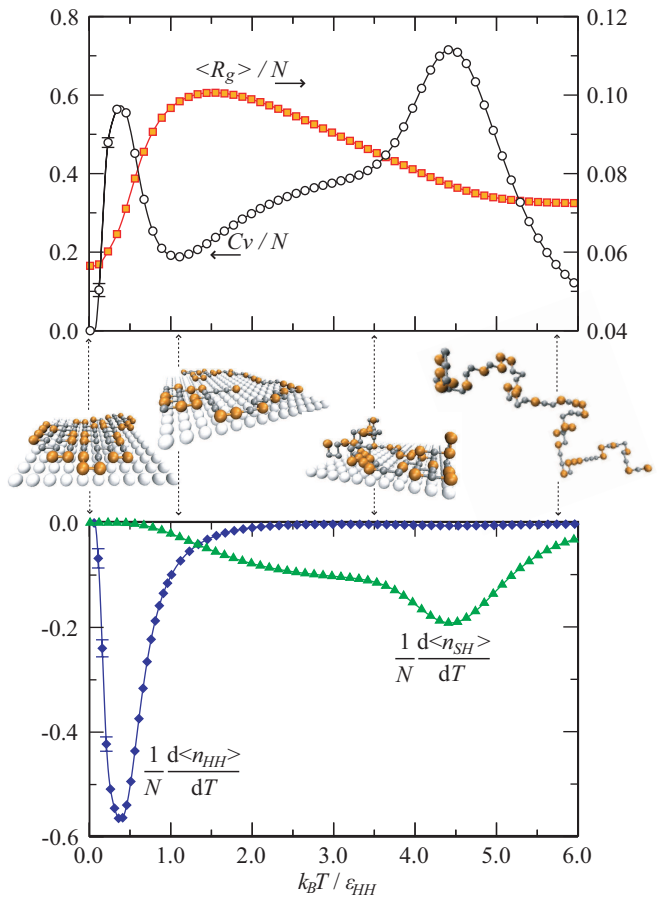}
	\vspace{0.5mm}
  \caption{(Color online) Thermodynamics of the 67mer interacting with surface
           $A$2 ($\varepsilon_{HH} = 1, \varepsilon_{SH} = \varepsilon_{SP}
           = 2$), which shows a typical category I transition.
           \textbf{Upper panel:} Specific heat, $C_V / N$, and the average radius of
           gyration per monomer, $\left\langle R_g \right\rangle / N$, as a function
					 of the effective temperature $k_BT / \varepsilon_{HH}$. The horizontal
					 arrows beside the labels indicate the axes to which the quantities refer.
           \textbf{Middle panel:} Typical configurations at different temperatures.
           \textbf{Lower panel:} Derivatives of the average numbers of H-H
           contacts per monomer, $(1/N)d\left\langle n_{HH} \right\rangle / dT$, and
           that of the average number of surface contacts of H monomers per monomer,
					 $(1/N)d\left\langle n_{SH} \right\rangle / dT$, as a function of
					 $k_BT / \varepsilon_{HH}$. Errors smaller than the data points are not
					 shown.}
  \label{thermoHP02HH01}
\end{figure}

Figure \ref{thermoHP02HH01} shows a typical transition pattern of category I demonstrated by
the 67mer with surface $A$2, for which $\varepsilon_{HH} = 1, \varepsilon_{SH}
= \varepsilon_{SP} = 2$. It is characterized by two pronounced peaks in $C_V$,
with $\left\langle R_g \right\rangle$ attaining its maximum between them as
seen in the upper panel of the figure. The nature of transitions to which the
two peaks in $C_V$ correspond can be identified by comparing them with $d\left\langle
n_{HH} \right\rangle / dT$ and $d\left\langle n_{SH} \right\rangle / dT$ in the
lower panel. Since the surface attracts both types of monomers equally,
$d\left\langle n_{SP} \right\rangle / dT$ shows similar behavior as $d\left\langle
n_{SH} \right\rangle / dT$ and thus is not shown in the figure. The $C_V$ peak
at $k_BT / \varepsilon_{HH} \approx 4.4$ represents the desorption-adsorption
transition where $d\left\langle n_{SH} \right\rangle / dT$ peaks at the same
temperature. The $C_V$ peak at $k_BT / \varepsilon_{HH} \approx 0.4$ represents the H-core
formation as $d\left\langle n_{HH} \right\rangle / dT$ also shows a peak at that
position. $\left\langle R_g \right\rangle$ decreases most rapidly during this latter process
when temperature is lowered.

A closer look at $C_V$ in Fig. \ref{thermoHP02HH01} shows a weak bump between
$k_BT / \varepsilon_{HH} \approx 1.0$ and $3.5$. The same phenomenon is also
observed in $d\left\langle n_{SH} \right\rangle / dT$ (and $d\left\langle n_{SP}
\right\rangle / dT$), suggesting that a subtle process related to the interaction
with the surface is taking place in this temperature range. Recalling our previous
discussion on the radius of gyration $\left\langle R_g \right\rangle$ in Section
\ref{limit}, this is a region where the HP chain keeps forming contacts with the
surface until it adsorbs completely on the surface. We call this process a
``flattening'' of the structure. When the surface attraction is sufficiently
strong, it occurs right after the chain is adsorbed to the surface but before
the H-core formation. The top graph in the leftmost column in Table
\ref{categories} shows a similar case for the 48mer, $\varepsilon_{HH} = 1,
\varepsilon_{SH} = \varepsilon_{SP} = 2$.

However, there are cases where this ``flattening'' bump is not observed in $C_V$,
as seen from the other two examples for the 67mer and 103mer with different surface
attractions in Table \ref{categories}. The flattening process might have been
``integrated'' within adsorption, or it simply does not cause enough energy fluctuations to give
a visible signal in $C_V$. In the latter case, signals can be found in other
structural quantities like $d\left\langle n_{SH} \right\rangle / dT$ or
$d\left\langle n_{SP} \right\rangle / dT$.


\subsubsection{Folding behavior with a moderately attractive surface: Category II}

\begin{figure}[th!]
  \includegraphics[width=\columnwidth]{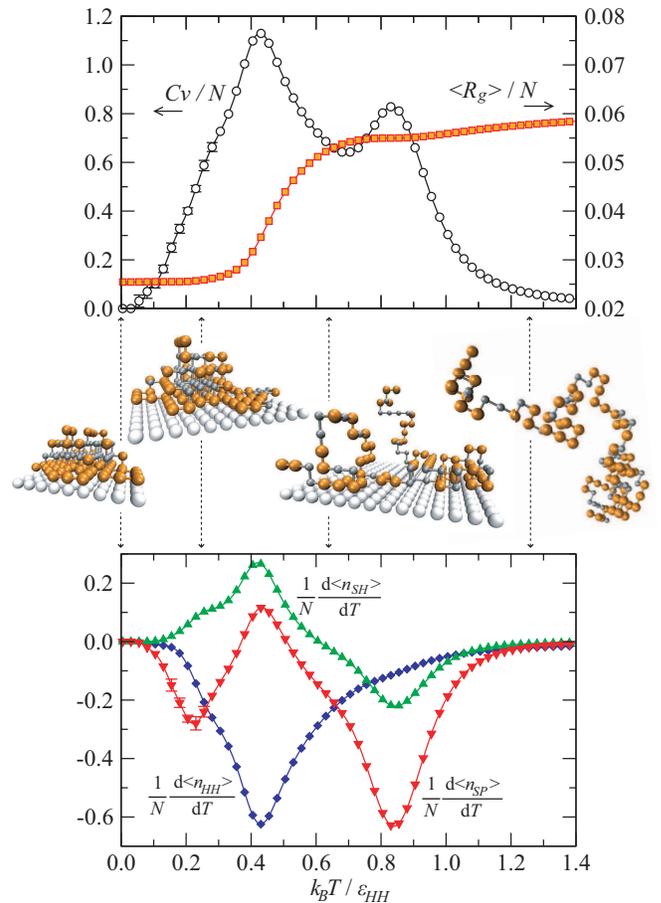}
	\vspace{0.5mm}
  \caption{(Color online) Thermodynamics of the 103mer interacting with surface
                 $P$\nicefrac{1}{2} ($\varepsilon_{HH} = 2, \varepsilon_{SH} = 0,
                 \varepsilon_{SP} = 1$), which shows a typical
                 category II transition.
                 \textbf{Upper panel:} Specific heat, $C_V / N$, and the average radius of
                 gyration per monomer, $\left\langle R_g \right\rangle / N$, as a function
		 of the effective temperature $k_BT / \varepsilon_{HH}$. The horizontal
 		 arrows beside the labels indicate the axes to which the quantities refer.
                 \textbf{Middle panel:} Typical configurations at different temperatures.
                 \textbf{Lower panel:} Derivatives of the average numbers of H-H
                 contacts per monomer, $(1/N)d\left\langle n_{HH} \right\rangle / dT$, and
                 those of the numbers of surface contacts, $(1/N)d\left\langle n_{SH}
		 \right\rangle / dT$ and $(1/N)d\left\langle n_{SP} \right\rangle / dT$,
		 as a function of $k_BT / \varepsilon_{HH}$, respectively. Errors smaller
		 than the data points are not shown.}
  \label{thermo103P01HH02}
\end{figure}

Figure \ref{thermo103P01HH02} shows the thermodynamics for the 103mer with surface
$P$\nicefrac{1}{2}, a typical case in category II. Similar to category I, systems in
category II also show two pronounced peaks in $C_V$ and identification of
structural transitions depends on the derivatives of $\left\langle n_{HH}
\right\rangle$, $\left\langle n_{SH} \right\rangle$ and $\left\langle n_{SP}
\right\rangle$. The peak at $k_BT / \varepsilon_{HH} \approx 0.85$ represents the
desorption-adsorption transition as identified by the peaks in $d\left\langle
n_{SH} \right\rangle / dT$ and $d\left\langle n_{SP} \right\rangle / dT$.
Another peak at $k_BT / \varepsilon_{HH} \approx 0.42$ indicates the H-core
formation as signaled by a peak in $d\left\langle n_{HH} \right\rangle / dT$.

The feature that differentiates category II from category I is the absence of
a maximum for $\left\langle R_g \right\rangle$ between the two peaks in $C_V$.
It decreases upon cooling until the very low temperature regime. The difference
arises from the fact that the flattening of structures occurs at a lower
temperature than the H-core formation in the vicinity of a less attractive
surface, giving rise to another transition hierarchy than category I. Two
possibilities for $\left\langle R_g \right\rangle$ are then observed when the
temperature is further lowered: (a) it keeps descending as in Fig.
\ref{thermo103P01HH02}; (b) it rises back up until $T = 0$, forming a minimum
below the H-core formation temperature as the 48mer does in Table \ref{categories}
(top graph, middle column for category II).

Interesting observations at low temperature are revealed by the thermodynamics of
$d\left\langle n_{HH} \right\rangle / dT$, $d\left\langle n_{SH} \right\rangle /
dT$ and $d\left\langle n_{SP} \right\rangle / dT$ as shown in the lower panel of
Fig. \ref{thermo103P01HH02}. During H-core formation at $k_BT / \varepsilon_{HH}
\approx 0.42$ where $d\left\langle n_{HH} \right\rangle / dT$ peaks at, troughs
are observed in $d\left\langle n_{SH} \right\rangle / dT$ and $d\left\langle
n_{SP} \right\rangle / dT$. This is a process of ``thickening'' during which
some of the surface attachments have to be broken to facilitate the construction
of H-H interactions.

When the temperature is further lowered to $k_BT / \varepsilon_{HH} \approx 0.25$,
a subtle shoulder could barely be seen in $C_V$ and $\left\langle R_g \right\rangle$
stays still on cooling; $d\left\langle n_{SP} \right\rangle / dT$, however, shows
a clear peak. This suggests that surface contacts for the P monomers are
established, demonstrating the flattening effect. Eventually the structures with
minimal possible energy are attained but they no longer span as many layers
vertically as at higher temperature. These structures are not completely planar
as in category I, as forming surface contacts is not always more energetically
favorable than forming hydrophobic H-H interactions.

In many other examples of this transition category (see Table \ref{categories}),
$C_V$ only has two major transition peaks, sometimes with a subtle shoulder or a
spike merged into the peaks as a result of a combination of various events.
Individual investigation of structural measures is thus essential to segregate
different structural changes.


\subsubsection{Folding behavior with a weak attractive surface: Category III and beyond}

\begin{figure}[tb!]
  \includegraphics[width=0.97\columnwidth]{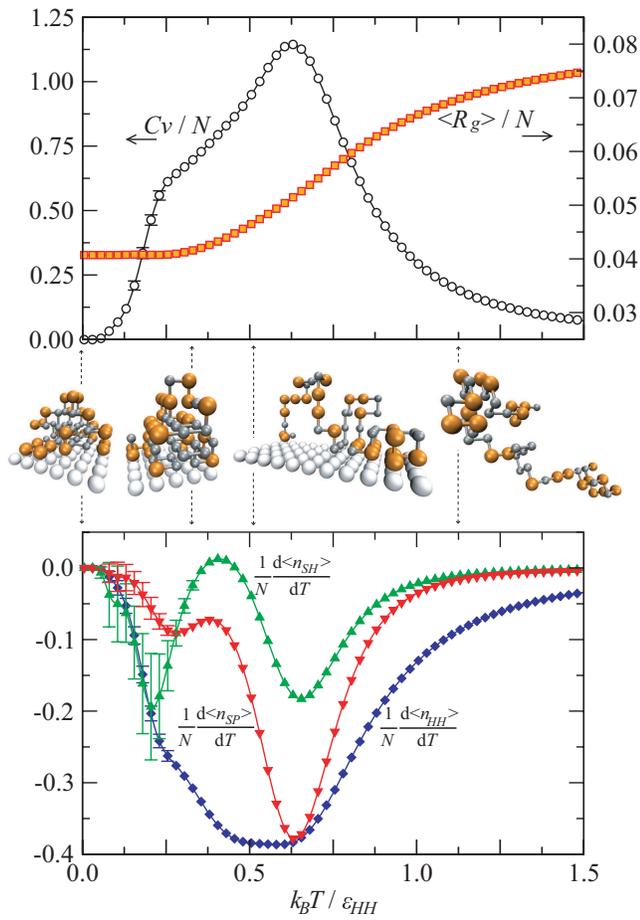}
	\vspace{0.5mm}
  \caption{(Color online) Thermodynamics of the 48mer interacting with surface
                 $P$\nicefrac{1}{2} ($\varepsilon_{HH} = 2, \varepsilon_{SH} = 0,
		 \varepsilon_{SP} = 1$), which shows a typical category III transition.
                 See Fig. \ref{thermo103P01HH02} for figure explanations.}
  \label{thermoP01HH02}
\end{figure}
When the surface attractive strength further reduces, the adsorption and
flattening temperatures decrease accordingly. Category III is identified when
the adsorption transition coincides with H-core formation, giving a single
peak in $C_V$ associated with a shoulder in some cases like the example shown in
Fig. \ref{thermoP01HH02}. The thermodynamics of category III transitions looks
similar to that in 3D free space. In both cases, $\left\langle R_g \right\rangle$ decreases upon
cooling and $C_V$ peaks at nearly the same temperature, except that
a higher peak results for systems of category III. Since adsorption and H-core formation
now occur almost together at nearby temperatures, more conformational degrees of
freedom are introduced by the surface interactions and this higher entropy gain
results in a larger $C_V$.

Details of transitions are again provided by $d\left\langle n_{HH}
\right\rangle / dT$, $d\left\langle n_{SH} \right\rangle / dT$ and
$d\left\langle n_{SP} \right\rangle / dT$. From the positions of peaks
in $d\left\langle n_{SH} \right \rangle / dT$ and $d\left\langle
n_{SP} \right\rangle / dT$, one may identify adsorption at $k_BT /
\varepsilon_{HH} \approx 0.63$ and flattening at $k_BT /
\varepsilon_{HH} \approx 0.24$ respectively. $d\left\langle n_{HH}
\right\rangle / dT$ manifests a wide peak across the adsorption and
flattening temperatures, which suggests that the hydrophobic core is
formed roughly in the temperature range $k_BT / \varepsilon_{HH}
\approx 0.25 - 0.87$. Instead of producing individual peaks in $C_V$,
the signals of adsorption and flattening are ``bridged'' and smoothed
out by the H-core formation, giving only a peak with a shoulder in
$C_V$, the shape and location of the latter being often subject to
some variability. Therefore, it is necessary to rely again on
structural quantities to separate signals for the various transitions
as illustrated in the previous categories.

For very weak attractive surfaces, adsorption and flattening occur at even lower
temperatures. They become distinguishable from the H-core formation which takes
place at a higher temperature, forming two or even three distinct peaks in $C_V$.
We generally classify systems with this transition hierarchy as category IV. For
a detailed discussion of an example from this category, see \cite{li2,wuest}.


\section{Discussion \label{discussion}}

\subsection{Remarks on the structural measures and categories}
\begin{table*}[!thbp]
\caption{Characteristic thermodynamics for Categories I, II, and III: Specific heat,
         $C_V / N$, and average radius of gyration per monomer, $\left\langle R_g
				 \right\rangle / N$, as a function of the effective temperature $k_BT /
				 \varepsilon_{HH}$. Errors smaller than the data points are not shown.
\label{categories}}
\onecolumngrid
\begin{ruledtabular}
\begin{tabular}{p{0.04\textwidth}||p{0.3\textwidth}||p{0.3\textwidth}||p{0.3\textwidth}}
 & \centering{\bf Category I} & \centering{\bf Category II} & \centering{\bf Category III} \tabularnewline \hline
 & \vspace{0.1mm} \centering{48mer, $\varepsilon_{HH} = 1, \varepsilon_{SH} = 2, \varepsilon_{SP} = 2$} \vspace{1mm}
 & \vspace{0.1mm} \centering{48mer, $\varepsilon_{HH} = 2, \varepsilon_{SH} = 1, \varepsilon_{SP} = 1$}
 & \vspace{0.1mm} \centering{36mer, $\varepsilon_{HH} = 3, \varepsilon_{SH} = 1, \varepsilon_{SP} = 1$} \tabularnewline

 & \centering{\includegraphics[width=0.28\textwidth]{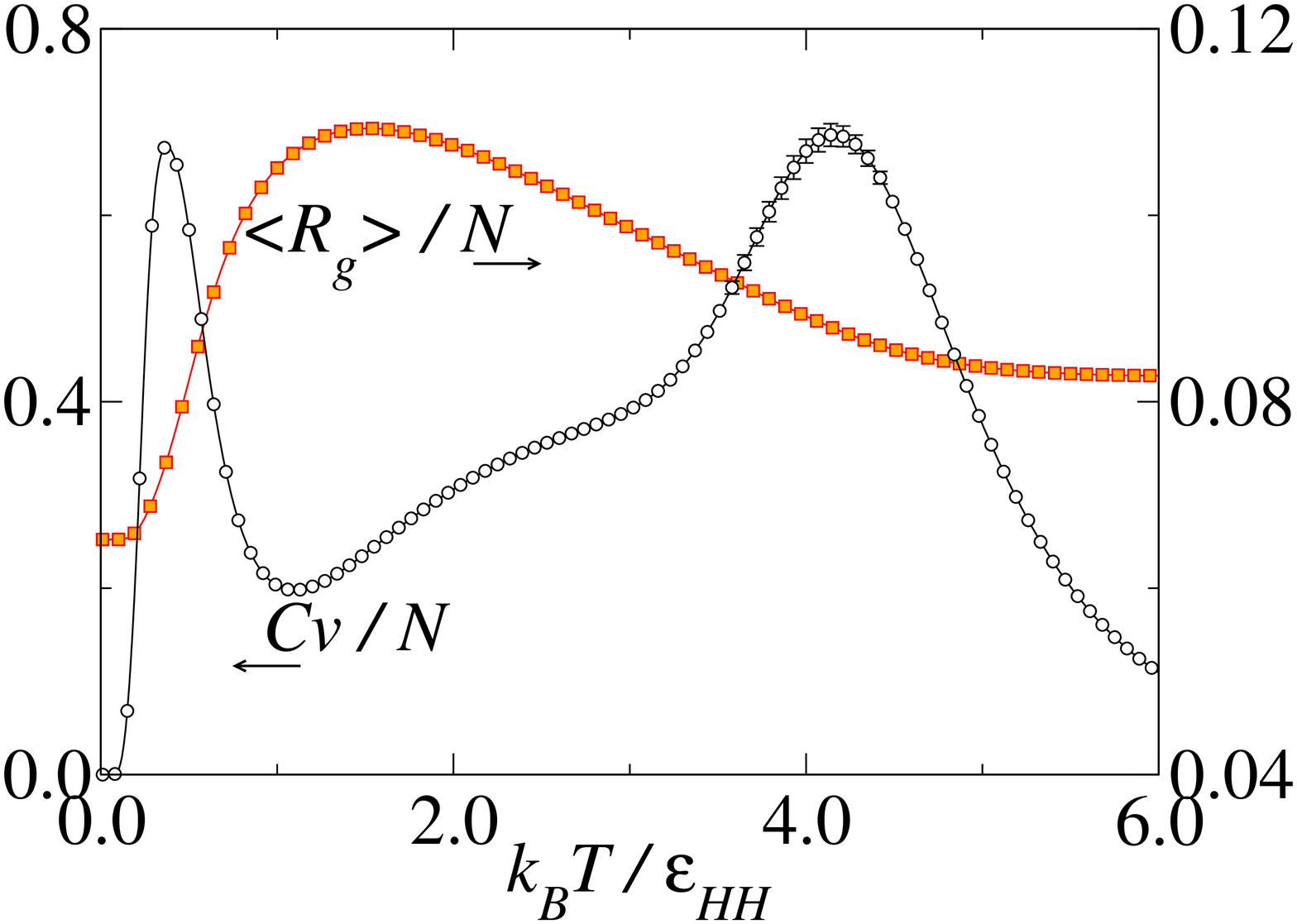}} \vspace{1mm}
 & \centering{\includegraphics[width=0.28\textwidth]{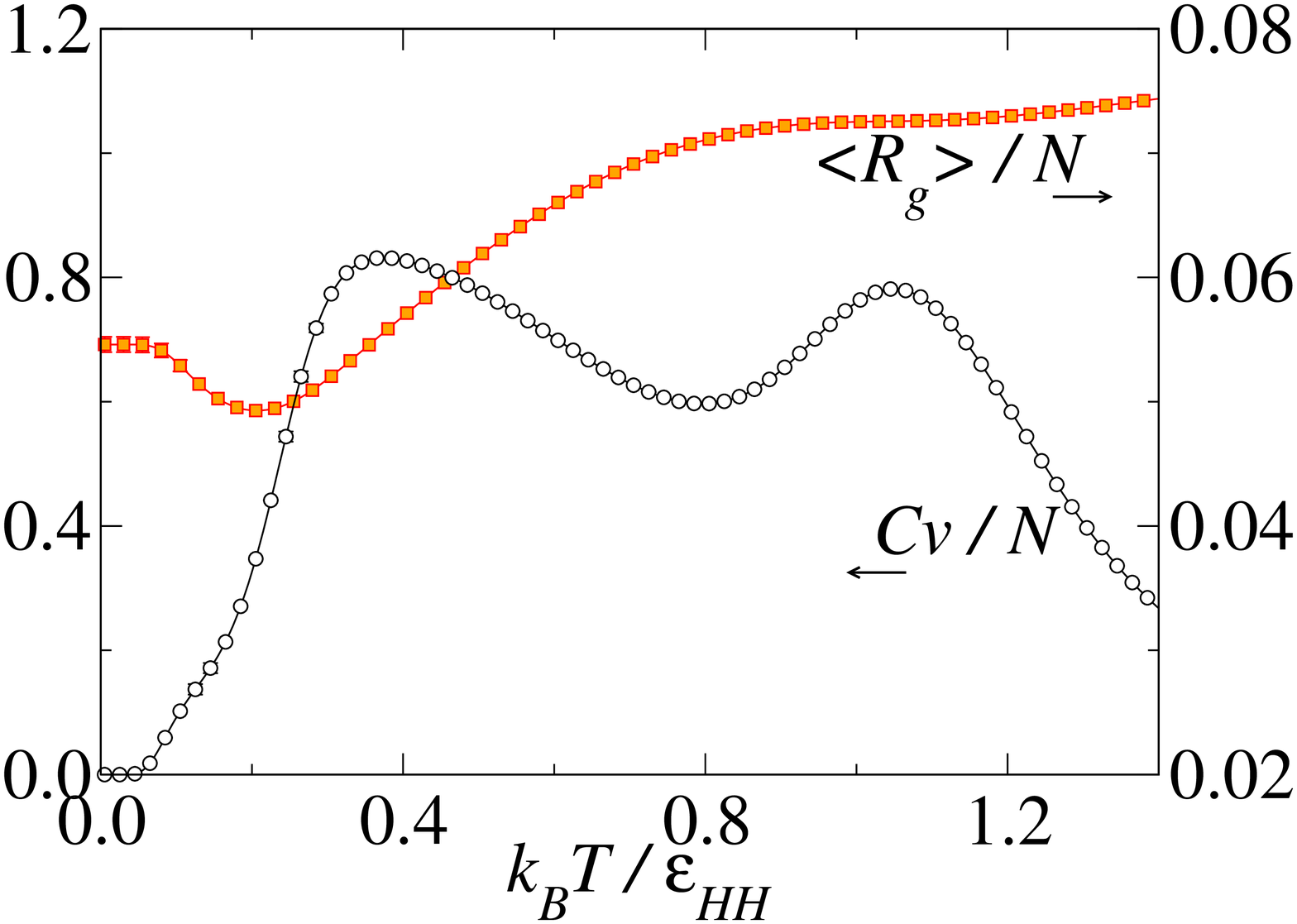}}
 & \centering{\includegraphics[width=0.28\textwidth]{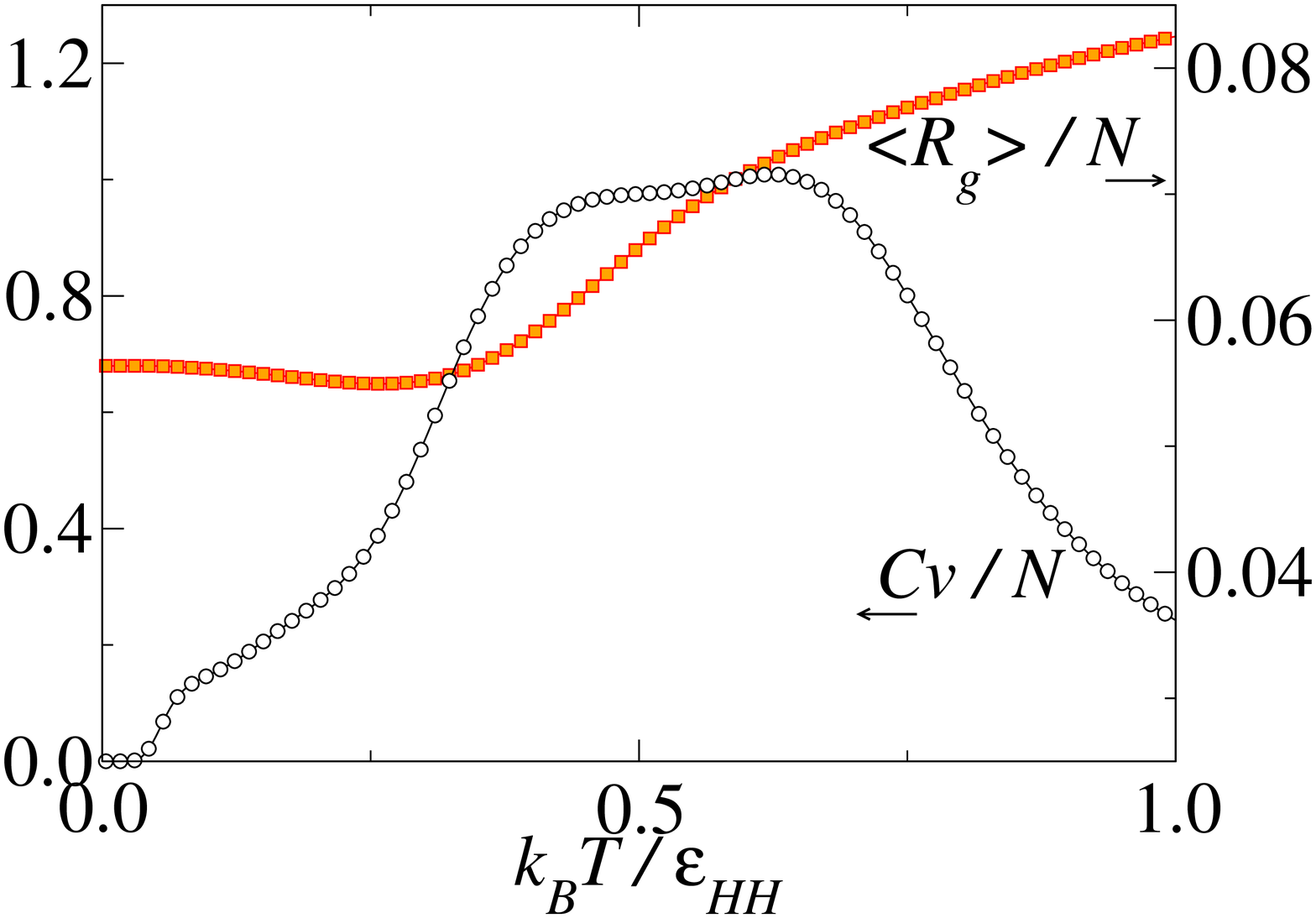}} \tabularnewline 


 & \vspace{0.1mm} \centering{67mer, $\varepsilon_{HH} = 1, \varepsilon_{SH} = 2, \varepsilon_{SP} = 0$} \vspace{1mm}
 & \vspace{0.1mm} \centering{67mer, $\varepsilon_{HH} = 1, \varepsilon_{SH} = 0, \varepsilon_{SP} = 1$}
 & \vspace{0.1mm} \centering{48mer, $\varepsilon_{HH} = 2, \varepsilon_{SH} = 1, \varepsilon_{SP} = 0$} \tabularnewline

 & \centering{\includegraphics[width=0.28\textwidth]{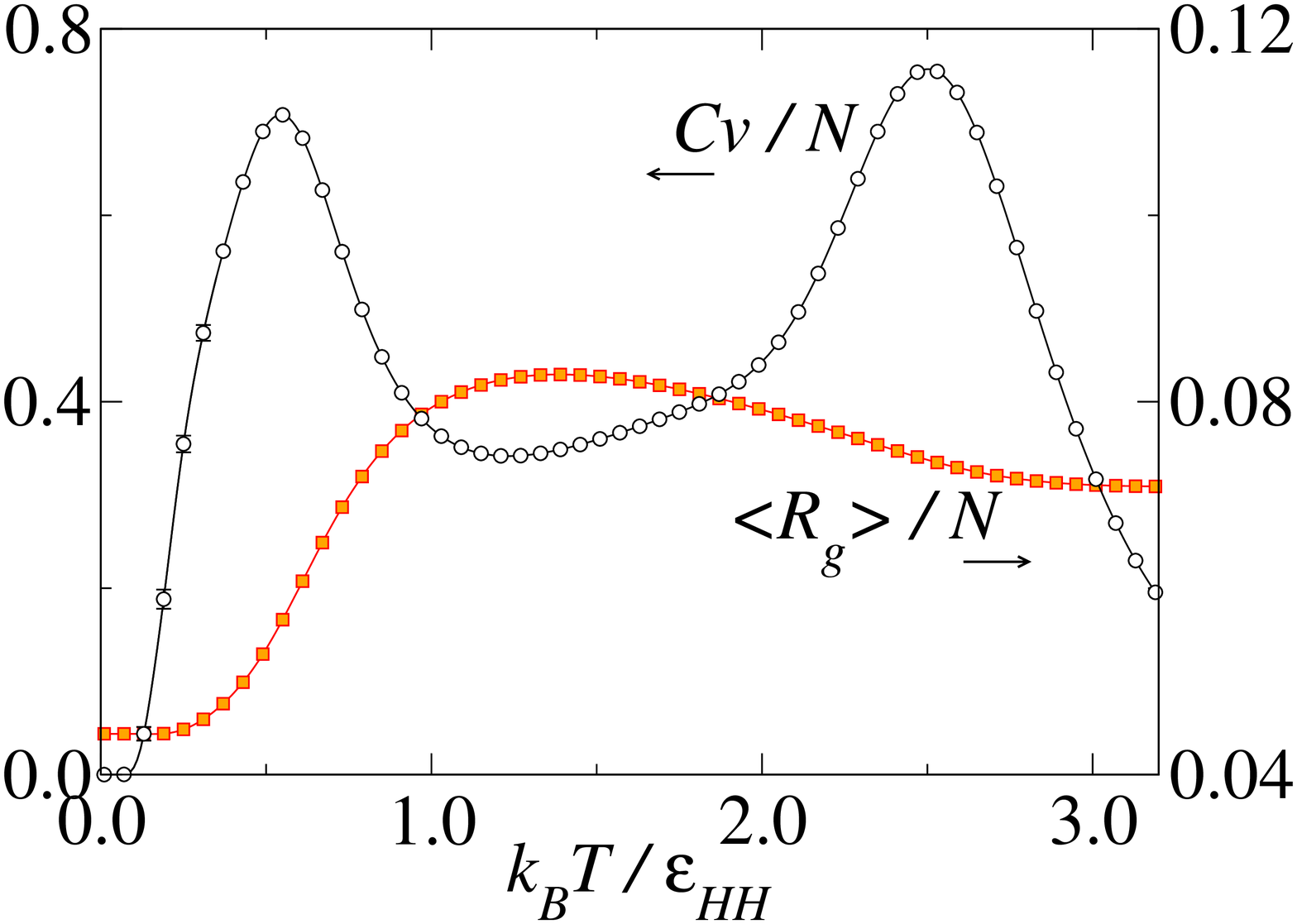}} \vspace{1mm}
 & \centering{\includegraphics[width=0.28\textwidth]{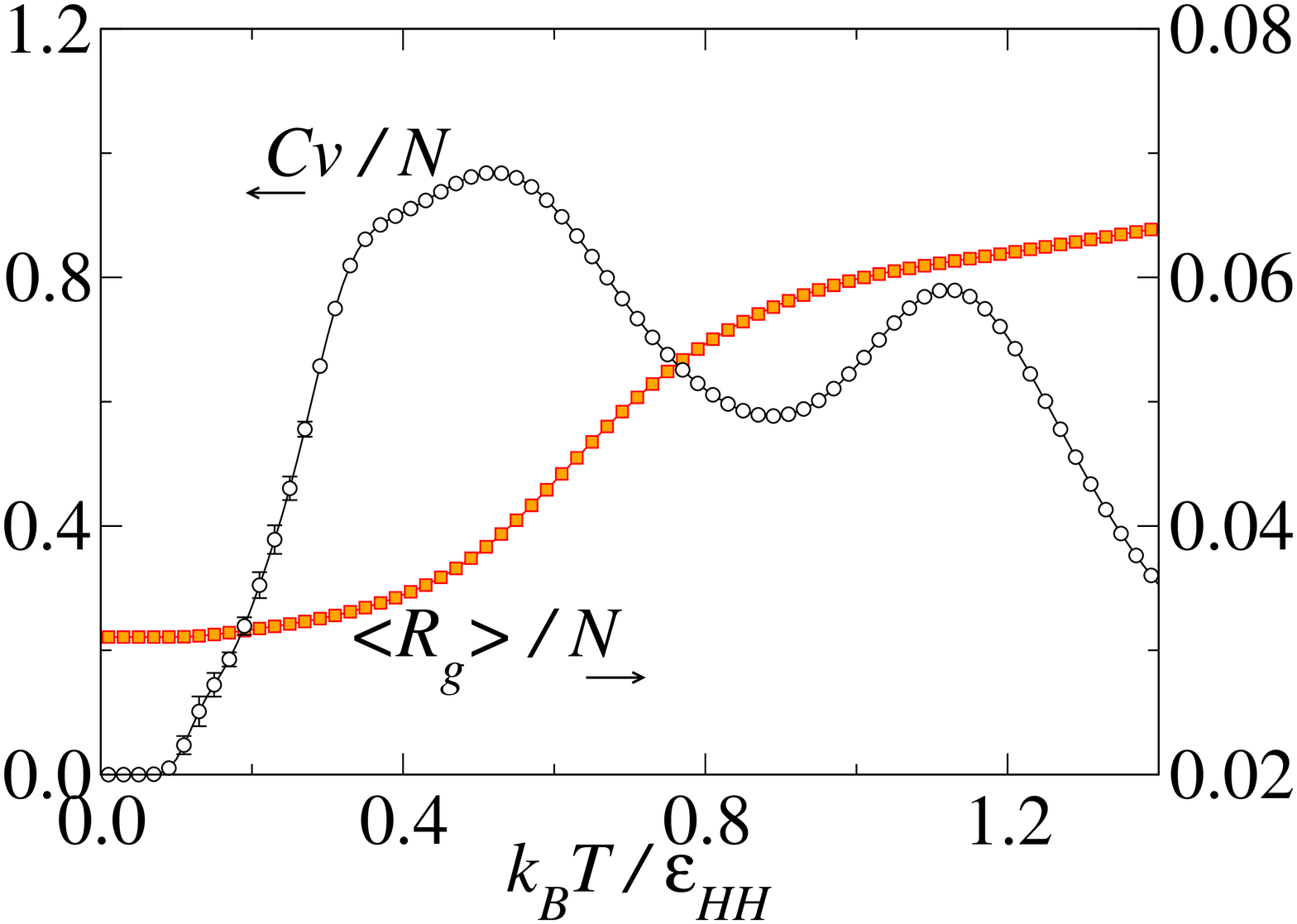}}
 & \centering{\includegraphics[width=0.28\textwidth]{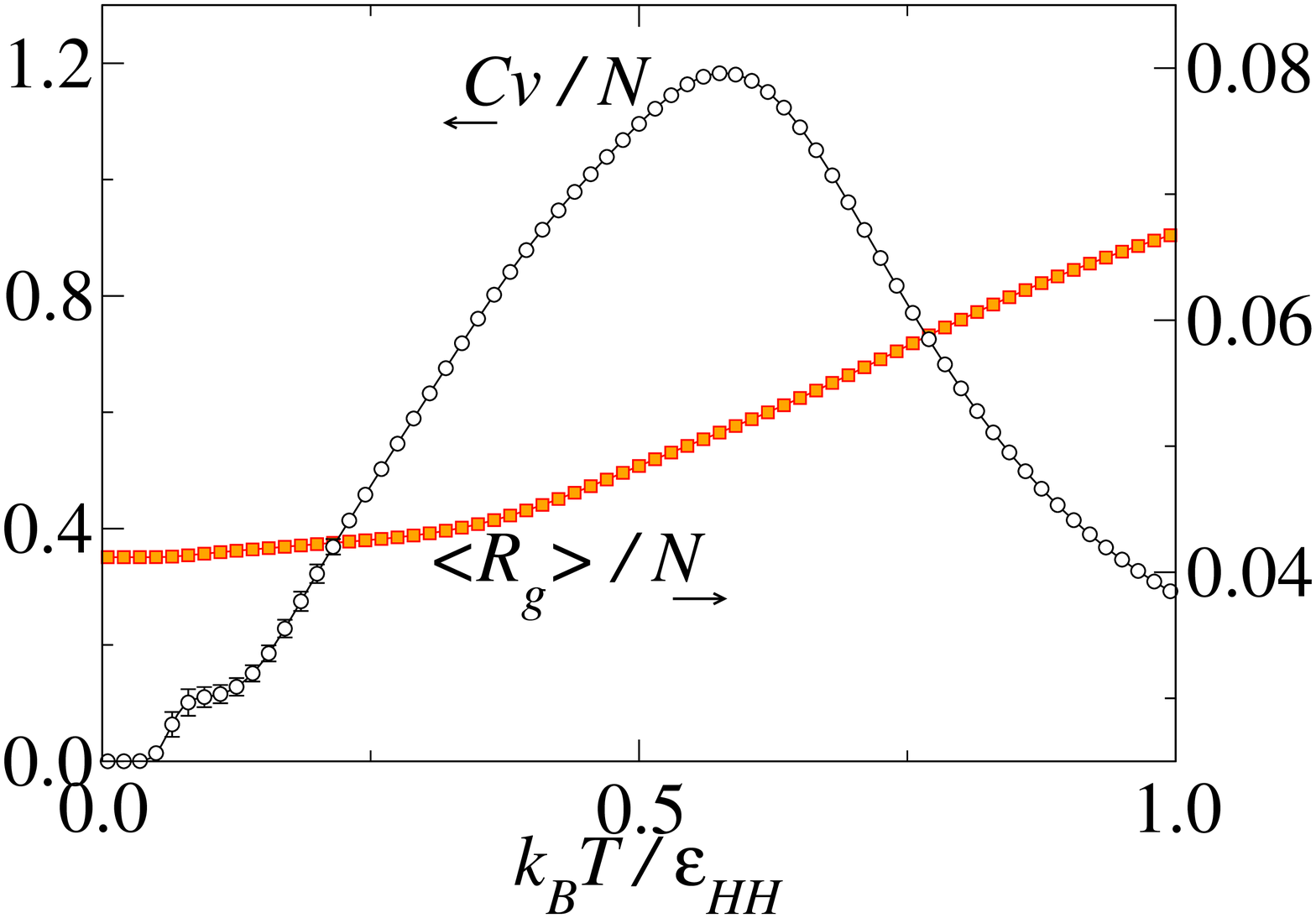}} \tabularnewline 

 & \vspace{0.1mm} \centering{103mer, $\varepsilon_{HH} = 1, \varepsilon_{SH} = 1, \varepsilon_{SP} = 1$} \vspace{1mm}
 & \vspace{0.1mm} \centering{103mer, $\varepsilon_{HH} = 1, \varepsilon_{SH} = 1, \varepsilon_{SP} = 0$}
 & \vspace{0.1mm} \centering{103mer, $\varepsilon_{HH} = 2, \varepsilon_{SH} = 1, \varepsilon_{SP} = 0$} \tabularnewline

 & \centering{\includegraphics[width=0.28\textwidth]{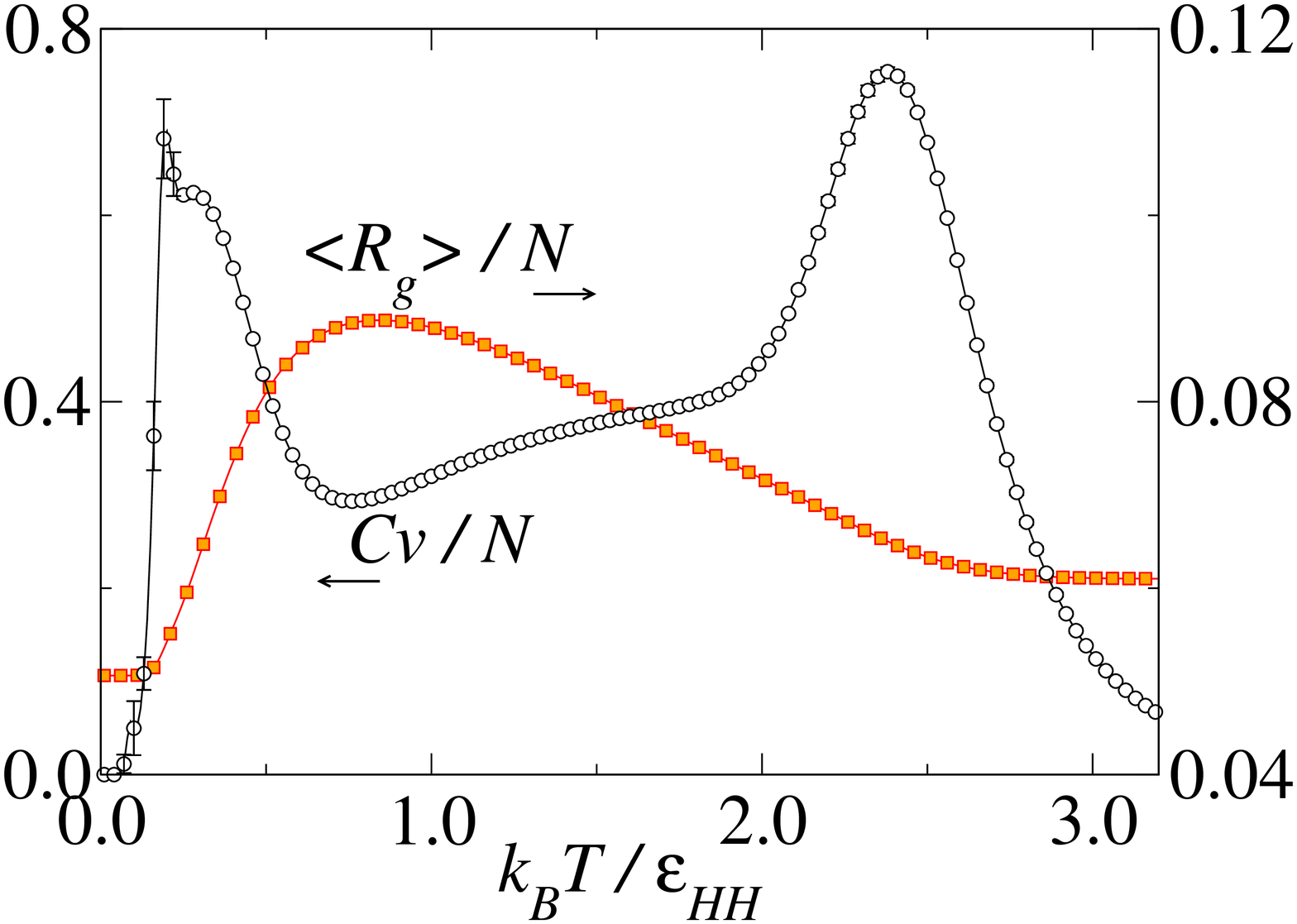}}  \vspace{1mm}
 & \centering{\includegraphics[width=0.28\textwidth]{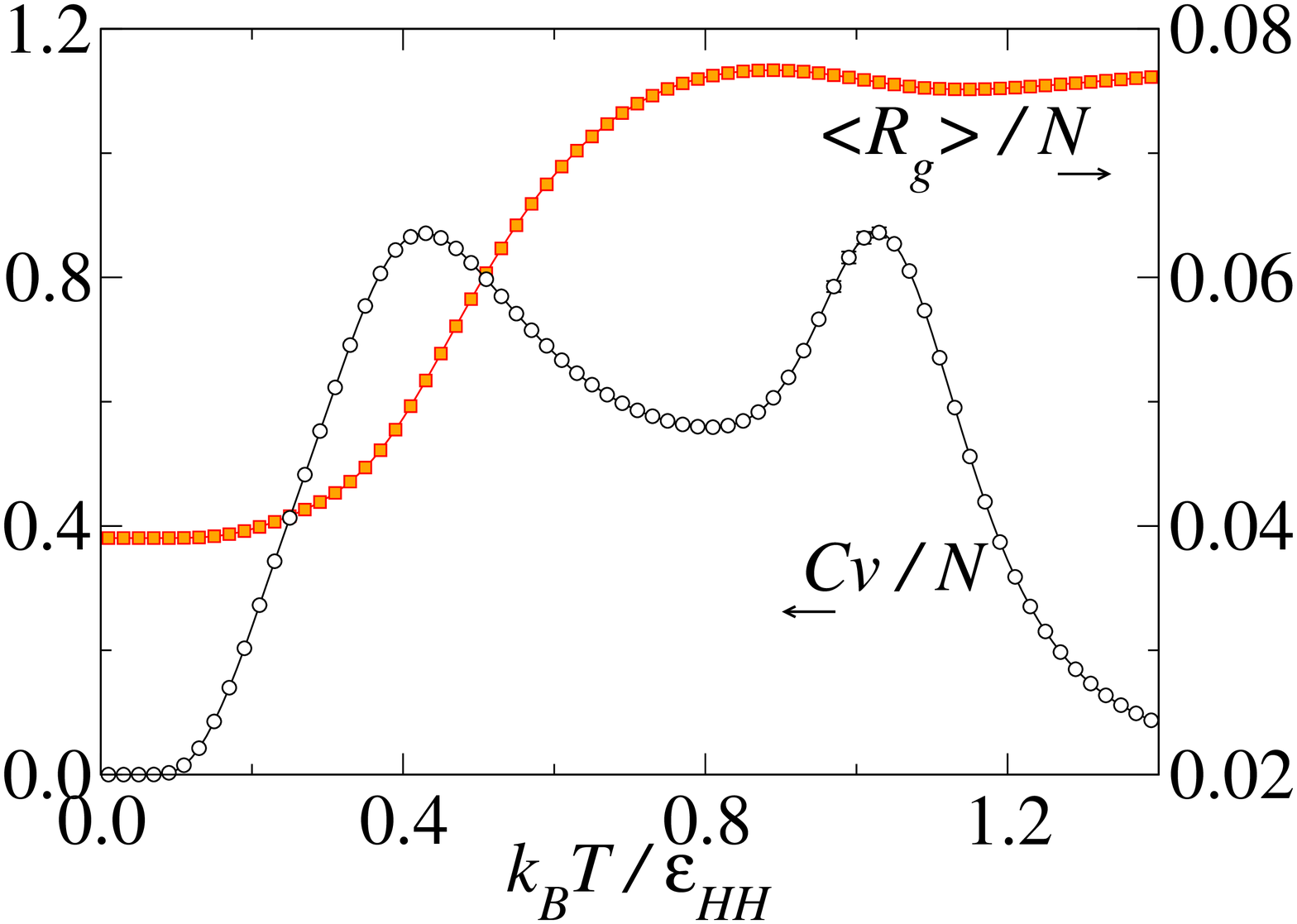}}
 & \centering{\includegraphics[width=0.28\textwidth]{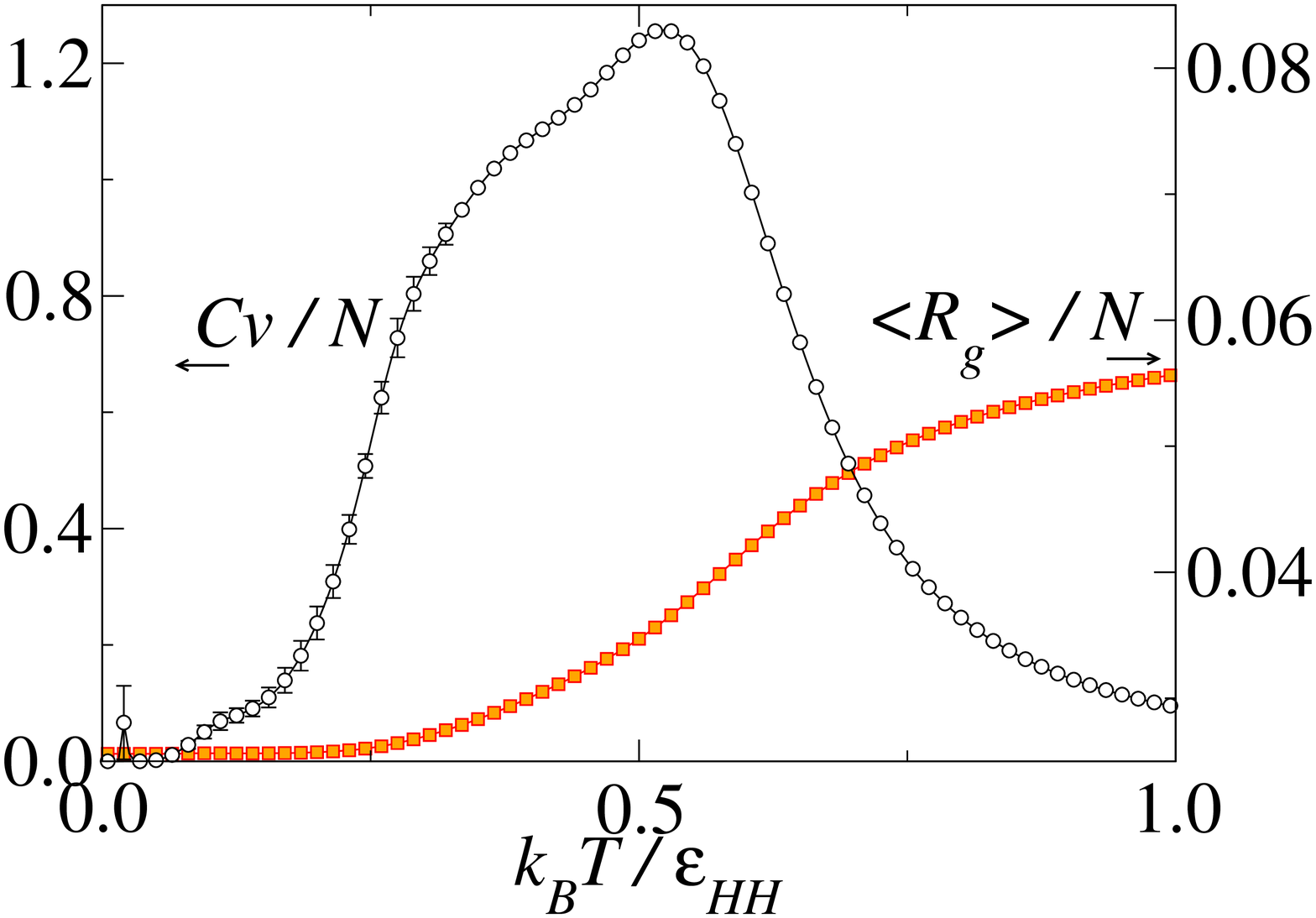}} \tabularnewline \hline

\multirow{8}{*}{$C_V$} &
  \begin{itemize}[leftmargin=3.5mm, rightmargin=2mm]
    \item shows two peaks, one for adsorption at high $T$, one for H-core formation at low $T$
    \item might show a bump for flattening between the two peaks
  \end{itemize}
&
  \begin{itemize}[leftmargin=3.5mm, rightmargin=2mm]
    \item shows two peaks, one for adsorption at high $T$, one for H-core formation at low $T$
    \item might have a shoulder or a spike for flattening below the H-core formation $T$
  \end{itemize}
&
  \begin{itemize}[leftmargin=3.5mm, rightmargin=2mm]
     \item shows only one peak for a combination of adsorption, flattening and H-core formation
     \item peak might have one or more shoulders for processes taking place at nearby $T$
  \end{itemize}
\tabularnewline \hline

\multirow{8}{*}{$\left\langle R_g \right\rangle$} &
  \begin{itemize}[leftmargin=3.5mm, rightmargin=2mm]
    \item shows a global maximum between the two $C_V$ peaks, signifying the end of flattening
          and the start of H-core formation upon cooling
  \end{itemize}
&
  \begin{itemize}[leftmargin=3.5mm, rightmargin=2mm]
    \item decreases upon cooling, might form a local maximum between the two $C_V$ peaks
    \item might rise back up due to flattening after H-core formation, forming a minimum
          at low $T$
  \end{itemize}
&
  \begin{itemize}[leftmargin=3.5mm, rightmargin=2mm]
    \item keeps decreasing upon cooling
    \item might rise back up due to flattening after H-core formation, forming a minimum
          at low $T$
  \end{itemize}
\tabularnewline

\end{tabular}
\end{ruledtabular}
\twocolumngrid
\end{table*}

Our results demonstrate that a comprehensive analysis of the specific
heat ($C_V$) combined with a set of appropriate structural quantities
is essential to shed light on recognizing structural transformations,
especially those subtle ones for which $C_V$ alone provides
insufficient information. We also note that in identifying phase
transitions, the peaks observed in structural quantities and those in
$C_V$ might be slightly off. One possible explanation could be finite
size effects \cite{sharma}. Nevertheless, this does not affect our
identification scheme much as these shifts are sufficiently small
compared to the difference in temperature scales required to clearly
identify distinct categories of transition patterns.

Table \ref{categories} further summarizes and compares the thermodynamic
features for the three categories discussed in the previous section. We have
intentionally chosen examples of various combinations of surface types (i.e.
surface strengths) and chain lengths for each category to illustrate that the
classification scheme is generic regardless of these variations.


\subsection{Classification of categories using relative surface attraction strengths}

Our classification scheme effectively generalized the folding behavior into
certain transition hierarchies. We further observe that the dominating factor
determining the transition category is closely associated with the relative
surface attraction, specifically, the ratio between $\varepsilon_{SH} +
\varepsilon_{SP}$ and $\varepsilon_{HH}$.

\begin{table}[th!]
\caption{Distribution of transition categories with respect to the relative
         surface attractions. The abbreviations refer to the surface types
	 introduced in Table \ref{energytable}. The numbers
         are the short forms of the benchmark sequences (e.g. 36 stands for
         the 36mer, 48.1 and 48.9 correspond to Seq. 1 and 9 among the ten
         48mers, respectively).
	 \label{histogram}}
\begin{ruledtabular}
\begin{tabular}{c|rlll|rlll|rlll}
$\varepsilon_{SH}$ & \multicolumn{4}{c|}{} & \multicolumn{4}{c|}{} & \multicolumn{4}{c}{} \tabularnewline
                +  & \multicolumn{4}{c|}{Category I} & \multicolumn{4}{c|}{Category II} & \multicolumn{4}{c}{Category III} \tabularnewline
$\varepsilon_{SP}$ & \multicolumn{4}{c|}{} & \multicolumn{4}{c|}{} & \multicolumn{4}{c}{} \tabularnewline \hline

\multirow{9}{*}{$ > \varepsilon_{HH}$} & $A$1: & 36,   & 48.1, & & & & & & & & & \\
                                       &     & 48.9, & 64,   & & & & & & & & & \\
                                       &     & 67,   & 103   & & & & & & & & & \\
                                       & $A$2: & 36,   & 48.9, & & & & & & & & & \\
                                       &     & 64,   & 67    & & & & & & & & & \\
                                       & $H$2: & 48.1, & 48.9, & & & & & & & & & \\
                                       &     & 64,   & 67    & & & & & & & & & \\
                                       & $P$2: & 48.1, & 48.9, & & & & & & & & & \\
                                       &     & 64,   & 67    & & & & & & & & & \\
\hline
\multirow{10}{*}{$ = \varepsilon_{HH}$} & \multicolumn{4}{r}{$A$\nicefrac{1}{2}:} & \multicolumn{4}{l|}{103} &&&& \\
                                        & \multicolumn{4}{r}{$P$1:}               & \multicolumn{4}{l|}{103} &&&& \\
                                        &                         &&&& $A$\nicefrac{1}{2}:  & 36,   & 48.1, &&&&& \\
                                        &                         &&&&                    & 48.9, & 64,   &&&&& \\
                                        &                         &&&&& 67    &       &&&&& \\
                                        &                         &&&& $H$1:                & 48.1, & 48.9, &&&&& \\
                                        &                         &&&&                    & 64,   & 67,   &&&&& \\
                                        &                         &&&&                    & 103   &       &&&&& \\
                                        &                         &&&& $P$1:                & 48.1, & 48.9, &&&&& \\
                                        &                         &&&&                    & 64,   & 67    &&&&& \\
\hline
\multirow{4}{*}{$ < \varepsilon_{HH}$} &&&&& $H$\nicefrac{1}{2}: & 67  &     && $H$\nicefrac{1}{2}: & 48.1, & 48.9, & \\
                                       &&&&&                   &     &     &&                   & 64,   & 103   & \\
                                       &&&&& $P$\nicefrac{1}{2}: & 67, & 103 && $P$\nicefrac{1}{2}: & 48.1, & 48.9  & \\
                                       &&&&&                   &     &     && $A$\nicefrac{1}{3}\footnote{$\varepsilon_{SH} = \varepsilon_{SP} = 1, \varepsilon_{HH} = 3$}: & 36, & 64 & \\
\end{tabular}
\end{ruledtabular}
\end{table}

Table \ref{histogram} shows the distribution of transition categories
against the relative surface attractions for systems with various
chain lengths and surface types. Ideally, a perfect correspondence
between the transition categories and the relative surface attractions
is implied if only the diagonal compartments were filled in Table
\ref{histogram}. In reality, as thermodynamic subtleties vary from
sequence to sequence, some off-diagonal compartments are also occupied
(e.g., some systems with $\varepsilon_{SH} + \varepsilon_{SP} <
\varepsilon_{HH}$ show category II behavior). A few systems also
reveal ``category duality'' where the thermodynamics bears properties
from both categories (e.g. the 103mer interacting with surfaces
$A$\nicefrac{1}{2} or $P$1). Nonetheless, the generality of our
classification scheme is clearly apparent and allows us to infer the
following basic rules: transition category I occurs for surfaces which
are strongly attractive ($\varepsilon_{SH} + \varepsilon_{SP} >\approx
\varepsilon_{HH}$); category II occurs when the hydrophobic internal
attraction is approximately comparable to the surface strengths
($\varepsilon_{SH} + \varepsilon_{SP} \approx \varepsilon_{HH}$);
category III can only occur when surface strengths are relatively weak
compared to $\varepsilon_{HH}$ ($\varepsilon_{SH} + \varepsilon_{SP} <
\varepsilon_{HH}$). Besides, we have also investigated
  the influence of the proportion of monomers by considering the ratio
  between $(n_{SH} / N) \varepsilon_{SH} + (n_{SP} / N)
  \varepsilon_{SP}$ and $\varepsilon_{HH}$. However, we did not find
  an obvious relation between these quantities and the transition
  categories. We thus believe that $\varepsilon_{SH} +
  \varepsilon_{SP}$ is more suitable for the classification scheme.

We stress that unlike other existing work, this classification is an
inference based on multiple HP sequences of various chain lengths and
attributes. In Table \ref{histogram}, we have also included results
from three other sequences, which were used as a ``testing set'' for
the adequacy of the classification scheme: a 36mer
(P$_3$H$_2$P$_2$H$_2$P$_5$H$_7$P$_2$H$_2$P$_4$H$_2$P$_2$HP$_2$);
another 48mer
(HP\-H$_2$\-P$_2$H$_4$PH$_3$P$_2$H$_2$P$_2$HPH$_3$PHPH$_2$P$_2$H$_2$P$_3$HP$_8$H$_2$);
and a 64mer
(H$_{12}$PHPHP$_2$H$_2$P$_2$H$_2$P$_2$HP$_2$H$_2$P$_2$H$_2$P$_2$HP$_2$\-H$_2$\-P$_2$H$_2$P$_2$HPHPH$_{12}$).
The 36mer and the 64mer were originally proposed to test a 2D genetic
algorithm \cite{unger}, whereas the 48mer is Seq. 1 of the ten testing
sequences in \cite{48mer}. All results from these additional sequences
fall into the diagonal compartments in Table~\ref{histogram},
reinforcing that our classification scheme is applicable to other
sequences interacting with an adsorbing substrate without loss of
generality. This is thus a breakthrough in our understanding of
adsorption properties of lattice proteins: Instead of
sequence-dependent individual behavior, the thermodynamics of HP
proteins do follow common patterns in structural transitions when they
interact with an adsorbing substrate.

Even if $\varepsilon_{SH}$ and $\varepsilon_{SP}$ have
the same magnitude, they are generally not expected to contribute
equivalently to the total energy because of the competition between
$\varepsilon_{SH}$ and $\varepsilon_{HH}$ for the H
monomers. However, the contrary appears to be the case in our study
because of the entropic effects, in which the adsorbing P monomers
also hinder the formation of H-H contacts in an indirect manner by
dragging the H monomers to the surface. Consider the surface $P$2 case
for instance: As it is energetically more favorable to form
surface-P contacts than internal H-H contacts, the chain tends to sit
on the surface rather than to form a compact globule. The H monomers
are then restricted to sit also on the surface instead of forming H-H
contacts, causing both $d \langle n_{SH} \rangle / dT$ and $d
\langle n_{SP} \rangle / dT$ to peak at the adsorption temperature.


\section{Conclusion \label{conclusion}}

In this work, protein adsorption has been studied with a
coarse-grained lattice model, the HP model, interacting with a surface
which either attracts all monomers, only hydrophobic monomers or only
polar monomers. We have employed Wang-Landau sampling with two
effective Monte Carlo trial moves, pull moves and bond-rebridging
moves, to obtain the energy density of states and subsequent
thermodynamics of structural quantities for a 48mer, 67mer and
103mer. Ground state energies are also reported. Based on the folding
and adsorption behavior revealed by a careful, comprehensive analysis
of the specific heat, radius of gyration and derivatives of the
numbers of surface contacts, we have been able to identify four main
types of transition hierarchies (three of which are discussed in
detail in this work). We have found that the occurrence of these
transition hierarchies is mainly determined by the attractive
couplings of the surface relative to the internal hydrophobic
attraction, i.e., the ratio between $\varepsilon_{SH} +
\varepsilon_{SP}$ and $\varepsilon_{HH}$, regardless of the surface
type, chain length or composition of H and P monomers of an HP
sequence.  Three other benchmark sequences, a 36mer, another 48mer and
a 64mer, have confirmed the validity of our classification scheme.

Although other transition categories cannot be excluded from our
study, the ones presented here provide a general and representative
picture of the thermodynamics of HP proteins interacting with an
adsorbing substrate. Our study also demonstrates that classifying
transition hierarchies by a combined analysis of the specific heat and
appropriate structural parameters provides a powerful route in
approaching similar systems of large conformational and sequence
spaces, for instance, HP proteins interacting with two confining,
attractive surfaces \cite{pattanasiri,pattanasiri2}.

However, further investigation is necessary to determine if there is a
rigorous relation between the proposed transition categories and the
relative surface attractions. More statistics from longer chains, or
chains of the same length but with different H and P compositions
would help clarifying the problem. The next question is whether the
same conclusions could be drawn, or what discrepancies would be found,
for other lattice models with other energy functions, i.e., different
interactions between monomers and with the surface. Another important
question is whether similar classification schemes could hold for
off-lattice models. In this case, thermodynamics of other structural
parameters, e.g., the gyration tensor, density profile, or any
suitable ones, could help in identifying such generic transition
patterns. All these together are essential in determining the
effectiveness of using different simplified protein models in computer
simulations to study protein adsorption from a macroscopic
perspective. \\ \\


\begin{acknowledgments}
We thank M. Bachmann for constructive discussions. This work was supported by
the National Science Foundation under grant no. DMR-0810223.
\end{acknowledgments}



\end{document}